# Cooperative effects in inelastic tunneling

M. Galperin[†] and Abraham Nitzan[††]

[†]Department of Chemistry & Biochemistry, University of California at San Diego, La Jolla CA 92093, USA

[††]Department of Chemistry, Tel Aviv University, Tel Aviv, 69978, Israel

## Abstract

Several aspects of intermolecular effects in molecular conduction have been studied in recent years. These experimental and theoretical studies, made on several setups of molecular conduction junctions, have focused on the current-voltage characteristic that is usually dominated by the elastic transmission properties of such junctions. In this paper we address cooperative intermolecular effects in the inelastic tunneling signal calculated for simple generic models of such systems. We find that peaks heights in the inelastic ($d^2I/dE^2$ vs. $E$) spectrum may be affected by such cooperative effects even when direct intermolecular interactions can be disregarded. This finding suggests that comparing experimental results to calculations made on single-molecule junctions should be done with care.

# 1. Introduction

The electronic conduction properties of molecular conduction junctions are determined by the electronic properties of the metal and molecular constituents, the bonding between them, the junction structure and configuration, external electrostatic (gate) fields and environmental parameters such as temperature. The number of molecules involved in the transmission process is another factor that may affect the junction conduction properties both qualitatively and quantitatively. Indeed, while some experiments show linear scaling of conduction with the number of involved molecules,[1-4] others show considerable deviations from such linear behavior,[5,6] and in particular Refs. 7 and 8 show marked cooperative effects that result from strong intermolecular interactions in conducting molecular islands. Qualitative understanding of such effects has been obtained from generic tight binding models and ab-initio simulations of junction comprising molecular islands or molecular layers,[9-14] as resulting from direct interactions between bridging molecules[15-17] [18] as well as indirect interactions mediated by the underlying substrate.[19-21] Other factors can play important roles in specific systems. Intermolecular interactions are particularly strong in polar molecular layers, [22-26] and such polarity may result from charge transfer to/from the substrate that itself may be affected by the adsorbate density.[24] Furthermore, the molecular bridge also affects, by its electrostatic screening behavior, the way an imposed voltage bias is distributed across the junction.[27,28]. Finally, junction structural response to the imposed field and consequently its electrical and thermal stability properties can reflect collective properties of its molecular and metal constituents.[29]

In this paper we address another, potentially important, aspect of such behavior in molecular conduction junctions - cooperative effects in inelastic tunneling. Inelastic electron tunneling spectroscopy (IETS) has become a principal diagnostic tool for the structure of such junctions but its applicability as such a tool rests to a large extent on the interpretation of observed signal in comparison to theoretical and computational studies.[30-38] The latter are often done in model junctions comprising a single bridge molecule. Clearly, cooperative effects in the inelastic signal can affect this interpretation, and their role and importance should be understood.

The paper is organized as follows. In Section 2 we introduce a simple generic model for describing cooperative molecular inelastic response in the conduction properties of molecular junctions. Section 3 outlines the calculation procedure, and shows to lowest order in the electron-vibration coupling how cooperative inelastic response can arise. Numerical results based on the exact mapping approach by Bonca and Trugman,[39,40] done on two- and three- molecules junctions are shown in Section 4, and some general observations are pointed out in the same Section. Section 5 concludes.

## 2. Model

We consider a model junction comprising two single-electron-level (two states, "empty" and "occupied") molecules, each with its own vibration and a standard polaron-type electron-vibration coupling, connected to two free electron reservoirs each in its own equilibrium. The corresponding Hamiltonian is

$$\hat{H} = \hat{H}_0 + \hat{V}^{(e-v)} + \hat{V}^{(et)} + \hat{V}^{(v-b)} \tag{1}$$

where

$$\hat{H}_0 = \sum_m \left( \varepsilon_m \hat{d}_m^\dagger \hat{d}_m + \omega_m \hat{b}_m^\dagger \hat{b}_m + \sum_\beta \omega_\beta \hat{b}_{m\beta}^\dagger \hat{b}_{m\beta} \right) + \sum_{k \in L,R} \varepsilon_k \hat{c}_k^\dagger \hat{c}_k \tag{2}$$

includes additively the independent electronic and vibrational degrees of freedom, while

$$\hat{V}^{(et)} = \sum_{K=L,R} \sum_{k \in K;m} \left( V_{km}^{(et)} \hat{c}_k^\dagger \hat{d}_m + V_{mk}^{(et)} \hat{d}_m^\dagger \hat{c}_k \right) \tag{3a}$$

$$\hat{V}^{(e-v)} = \sum_m V_m^{(e-v)} \hat{Q}_m \hat{d}_m^\dagger \hat{d}_m \tag{3b}$$

$$\hat{V}^{(v-b)} = \sum_m \sum_\beta U_{m\beta}^{(v-b)} \hat{Q}_m \hat{Q}_{m\beta} \tag{3c}$$

describe interactions between them. Here ($et$) denotes the coupling associated with electron transfer between molecule and leads, ($e-v$) denotes the interaction between the tunneling electrons and the molecular vibrations and ($v-b$) is the coupling between the molecular vibrations and their corresponding (harmonic) thermal baths. $\hat{d}_m^\dagger$ ($\hat{d}_m$) and $\hat{c}_k^\dagger$ ($\hat{c}_k$) create (annihilate) an electron in the molecular state $m$ and in the lead state $k$ of energies $\varepsilon_m$ and $\varepsilon_k$, respectively. $\hat{b}_m^\dagger$ ($\hat{b}_m$) and $\hat{b}_{m\beta}^\dagger$ ($\hat{b}_{m\beta}$) create (annihilate) vibrational quanta in the molecular mode $m$ and the thermal bath mode, $m\beta$, respectively. Finally

$$\hat{Q}_j \equiv \hat{b}_j + \hat{b}_j^\dagger \qquad j = m, (m\beta) \qquad (4)$$

are displacement operators for the molecular ($m$) and the corresponding thermal bath ($m\beta$) vibrations of frequencies $\omega_m$ and $\omega_\beta$, respectively. Note we have represented the molecular system as a set of single electron levels, each coupled to its own vibration, which in turn is coupled to its own thermal harmonic bath. Note that in our model, cooperative transport effects result from the effective coupling between molecules due to their interaction with contacts - no direct inter-molecular coupling is assumed.[41] Seeing cooperative inelastic tunneling effects in this restrictive model will be an analog of the cooperative elastic tunneling discussed in Refs. 9-14, while stronger cooperative behavior may be expected if all modes are coupled to the same thermal bath. Furthermore, we consider tunneling through a systems of identical molecules and therefore take, in the calculations described below, the parameters $\varepsilon_m = \varepsilon_0$, $\omega_m = \omega_0$, $V_m^{(e-v)} = M_0$, $V_{km}^{(et)}$ and $U_{m\beta}^{(v-b)}$ independent of the molecular index $m$. (In the calculation described below the effect of the molecule-lead coupling $V_{km}^{(et)}$ is represented by local coupling parameters $t_0^L$ and $t_0^R$).

## 3. Calculation procedure

We are interested in the possibility that coherence in the subspace of the primary vibrational modes is induced by their coupling to the electronic subsystem. For this reason we avoid the methodology used in earlier works[42,43] where a mean field type approximation is used to factorize vibronic Green functions into their electronic and vibrational components. Instead, our calculation is done within the Bonca-Trugman framework[39,40] that uses an exact mapping of the many-body electron-vibration problem to a single electron scattering in the multidimensional vibrational state-space. The electron scattering is considered on a tight-binding chain in the space of vibronic states of the molecule. Figure 1 shows the scattering process for a system involving a one-electronic state and one vibrational mode impurity (bridge) connecting two leads that are represented by tight binding chains with nearest neighbor coupling (black lines connecting the sites), described in the extended space with $n_v$ vibrational levels (electron-vibrational coupling represented by green lines connecting sites on the impurity (blue) section). The red sites correspond to the incoming channel, where the electron approaches the impurity site (blue) from the left while the vibration is in the ground state. The scattering process couples this channel to the outgoing channels (black sites) defined for each of the $n_v$ vibrational levels. These outgoing channels enter the calculations via self-energy terms defined on the sites next to the impurity (cyan), and make it possible to evaluate the outgoing current $I^{\{v\}}$ associated with different final vibrational states $v$. In the language used to describe calculations of transmission in molecular conduction junctions, the blue and cyan sites in Fig. 1 constitute the "extended molecule" represented in the extended vibrational state space of the impurity. As described, the model of Figure 1 just represents the Bonca-Trugman problem. The new element in our calculation is that this model is extended a molecular impurity that comprises *N* electronic levels each coupled to its own vibration. (The reader may imagine the corresponding model by overlaying Fig 1 with *N* similar layers in the 3$^{rd}$ dimension, keeping in mind that the different outgoing channels are different from each other only in the state of their vibrational (*N* modes) subspace.

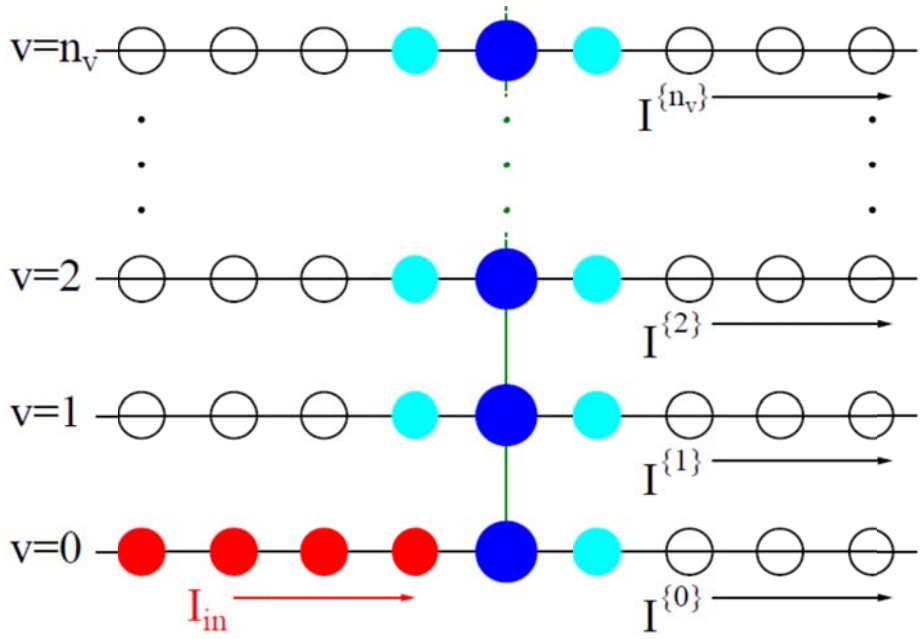

Fig. 1 Electron scattering in 1-dimension, in an extended vibrational space, for an impurity scatterer comprising one electronic level coupled to one vibrational mode. Each horizontal chain represents the 1-dimensional electron transport process, where the central site (blue) is the impurity and the sites on its two sides represent the leads. The chain replicas correspond to different vibrational states of the oscillator. The electron-vibration interaction is localized on the impurity, and is indicated by the vertical lines connecting the blue sites. See text for more details.

It should be emphasized that the calculation described below is a purely scattering calculation. The transmission coefficient as a function of energy of incoming electron T(E) represents conductance in the same sense as in the Landauer approach to transport in junctions. Since the incoming state is on the ground vibrational level, inelastic scattering corresponds to an outgoing state on a higher level, and its energy threshold is $E_{in} = \hbar\omega$, where $E_{in}$ is measured from the bottom of the electronic band. Note that because we use the outgoing vibrational energy as an indicator, vibrational relaxation in the outgoing channel due to coupling to the thermal environment is disregarded and coupling to this environment (assumed to be at zero temperature) is used only in taking the incoming state to correspond to the ground state vibration. Plotting the second derivative of the outgoing current with respect to $E_{in}$ will show a peak at this threshold in much the same way that such peaks are seen in inelastic tunneling spectroscopy. The dependence of this signal on the number $N$ of molecules in the junction can be used as an indicator to coherent effects in the inelastic transmission process.[44]

Some details of the calculation are described next. After the mapping the Hamiltonian is

$$\hat{H} = \sum_{\{v\}} \left[ \sum_{\ell=-\infty}^{-1} \left( \varepsilon^{\{v\}} \left| \psi_\ell^{\{v\}} \right\rangle \left\langle \psi_\ell^{\{v\}} \right| - t \left| \psi_\ell^{\{v\}} \right\rangle \left\langle \psi_{\ell-1}^{\{v\}} \right| + H.c. \right) \right.$$

$$+ \sum_{m=1}^{N} \left( -t_0^L \left| \psi_{-1}^{\{v\}} \right\rangle \left\langle \psi_m^{\{v\}} \right| + H.c. + \varepsilon_0^{\{v\}} \left| \psi_m^{\{v\}} \right\rangle \left\langle \psi_m^{\{v\}} \right| \right.$$

$$+ M(\{v\},\{v'\}) \left| \psi_m^{\{v\}} \right\rangle \left\langle \psi_m^{\{v'\}} \right| + H.c. - t_0^R \left| \psi_m^{\{v\}} \right\rangle \left\langle \psi_1^{\{v\}} \right| + H.c. \right)$$

$$\left. + \sum_{r=1}^{+\infty} \left( \varepsilon^{\{v\}} \left| \psi_r^{\{v\}} \right\rangle \left\langle \psi_r^{\{v\}} \right| - t \left| \psi_r^{\{v\}} \right\rangle \left\langle \psi_{r+1}^{\{v\}} \right| + H.c. \right) \right] \quad (5)$$

where $\{v\} \equiv \{v_1, \ldots, v_N\}$ is vibronic channel characterized by set of states of molecular vibrations, $M(\{v\},\{v'\}) \equiv \left\langle \psi_m^{\{v\}} \left| \hat{V}^{(e-v)} \right| \psi_m^{\{v'\}} \right\rangle$ are transition matrix elements between two vibrational channels defined by usual rules, and $\varepsilon^{\{v\}} \equiv \varepsilon + \omega_0 \sum_{m=1}^{N} v_m$ and $\varepsilon_0^{\{v\}} \equiv \varepsilon_0 + \omega_0 \sum_{m=1}^{N} v_m$ are vibronic energies in the contacts and the molecules, respectively. Finally, $t$, $t_0^L$ and $t_0^R$ are the nearest neighbor interstate coupling in the leads and the corresponding couplings between the molecule and the left and right leads, respectively.

As already noted, we restrict consideration to scattering of an electron of energy $E$ incoming in the vibrational ground state $\{v_0\} \equiv \{0,\ldots,0\}$ from the left contact. Its wave vector is $k_{in} = \arccos \dfrac{\varepsilon - E}{2t}$, and the incoming flux is $I_{in} = 2t \sin k_{in}$. Following the procedure outlined in Ref. 39,40 one derives for the reflection coefficient in the incoming channel $\{v_0\}$

$$r^{\{v_0\}} = \frac{te^{-ik_{in}} - AB}{te^{ik_{in}} - A^*B}; \quad A \equiv \frac{\varepsilon^{\{v_0\}} - E - te^{-ik_{in}}}{t} e^{-ik_{in}}; \quad B \equiv \varepsilon^{\{v_0\}} - E - \sum_{m,m'=1}^{N} \left(t_0^L\right)^2 R_{(\{v_0\}m)(\{v_0\}m')}$$

(6)

Here $R_{(\{v_0\}m)(\{v_0\}m')}$ is matrix element between vibronic states $(\{v_0\}m)$ and $(\{v_0\}m')$ of the resolvent (retarded Green function) in the molecular subspace. The latter includes all molecular sites as well as sites -1 (except $|\psi_{-1}^{\{v_0\}}\rangle$) and +1 from the left and right contacts, respectively, together constituting the "extended molecule". The rest of the contacts is accounted for by introducing self-energies

$$\varepsilon_s^{\{v\}} \to \tilde{\varepsilon}_s^{\{v\}} \equiv \varepsilon_s^{\{v\}} - \frac{t^2}{\varepsilon^{\{v\}} - E - te^{ik^{\{v\}}}}, \tag{7}$$

where $s = -1, 1$ and $k^{\{v\}} \equiv \arccos \frac{\varepsilon^{\{v\}} - E}{2t}$. Utilizing the form of the Schrodinger equation on a tight-binding chain in the left contact and expression (6) for the reflection coefficient yields

$$|\psi_{-1}^{\{v_0\}}\rangle = A + A^* r^{\{v_0\}} \text{ and } |\psi_m^{\{v\}}\rangle = \sum_{m'=1}^{N} R_{(\{v\}m)(\{v_0\}m')} t_0^L |\psi_{-1}^{\{v_0\}}\rangle.$$

The transmitted fluxes in vibrational channels on the right side of the junction

$$I^{\{v\}} = -2t \, \text{Im}\left[\langle \psi_2^{\{v\}} | \psi_1^{\{v\}} \rangle\right], \tag{8}$$

where

$$|\psi_2^{\{v\}}\rangle = \frac{t}{\varepsilon^{\{v\}} - E - te^{ik^{\{v\}}}} |\psi_1^{\{v\}}\rangle, \tag{9}$$

define the transmission coefficients for particular scattering channels $T^{\{v\}} = I^{\{v\}} / I_{in}$ ( $T^{elast} \equiv T^{\{v_0\}}$ is the elastic transmission probability and $T = \sum_{\{v\}} T^{\{v\}}$ is the total transmission coefficient). These coefficients represent channel-resolved and total conductance of the junction. Below we are interested in the off-resonant IETS signal, thus we plot derivative of the inelastic transmission coefficient

$$T^{inel} = T - T^{elast} = \sum_{\{v\} \neq \{v_0\}} T^{\{v\}} \tag{10}$$

with respect to energy of the incoming electron.

## 4. Results

Figures 2-5 show the results of our calculation. Unless otherwise stated, the parameters used are $\varepsilon = 1$, $t = 1$, $t_0^L = t_0^R \equiv t_0 = 0.5$, $\omega_0 = 1$, $M_0 = 1$ ($m \in \{1,\ldots,n\}$), $\bar{\varepsilon}_0 \equiv \varepsilon_0 - \dfrac{M_0^2}{\omega_0} = 3$. With these parameters, the energy band in the leads lies in the range $-2 \leq E \leq +2$ and the imaginary part of the self-energy at the band center is $\Gamma = -2\,\mathrm{Im}\left(\Sigma^<(E=0)\right) = 0.5$ ($\Sigma^<$ is the retarded self energy. Note that this choice of parameters puts the molecular resonance outside the bands of the chains representing the leads, but this causes no calculational or conceptual difficulty for a model representing an off resonance tunneling situation. (The left side of Figure 4 does show the behavior of our coherence measure as the molecular resonance enters the band at $\bar{\varepsilon}_0 < 2$). The inelastic tunneling threshold appears at tunneling energy $E_{threshold} = E_{bot} + \omega_0$, where $E_{bot} = -2t$ is the bottom of the leads' band.

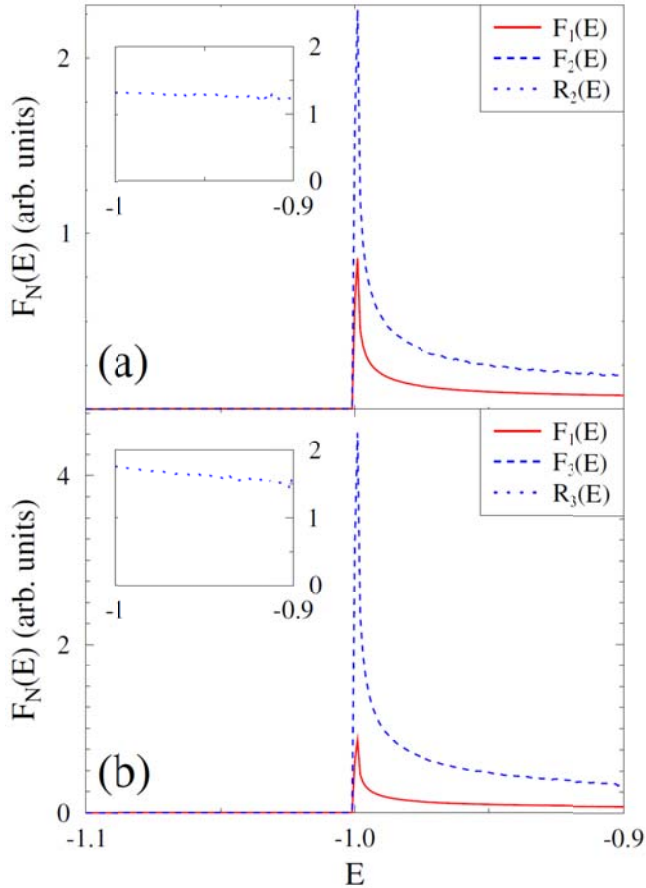

Fig. 2. The inelastic tunneling signal $d^2I/dE^2$, plotted vs. the tunneling energy $E$ in the vicinity of the energy threshold $E_{threshold}$ for inelastic tunneling, shown for molecular bridges with $N=1,2$ molecules in panel (a), and $N=1,3$ molecules in panel (b). The insets show the ratios $R_N(E) = F_N(E)/NF_1(E)$. See text for parameters.

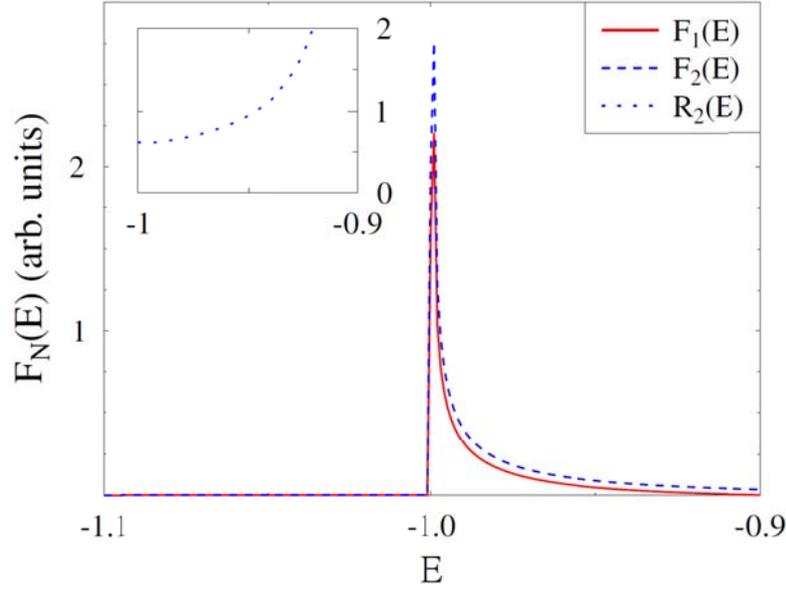

Fig. 3. Same as Fig 2a, with the same parameters (see text) except that $\varepsilon = 0.2$.

Figures 2a and 2b show, respectively, the inelastic tunneling peak in the standard $F_N(E) = \left(d^2I/dE^2\right)_N$ spectra, obtained for $N=2$ and $N=3$ molecules junctions together with the single molecule result. The insets in these figures show, for $E > E_{threshold}$, the ratios $R_N(E) = F_N(E)/NF_1(E)$. ($F_N(E) = 0$ for $E < E_{threshold}$). The deviation of this ratio from 1 indicates molecular cooperative behavior in the inelastic tunneling signal. While these results show constructive interference ($R > 1$), destructive interference ($R < 1$) can be obtained with other choices of parameters as seen in Fig. 3.

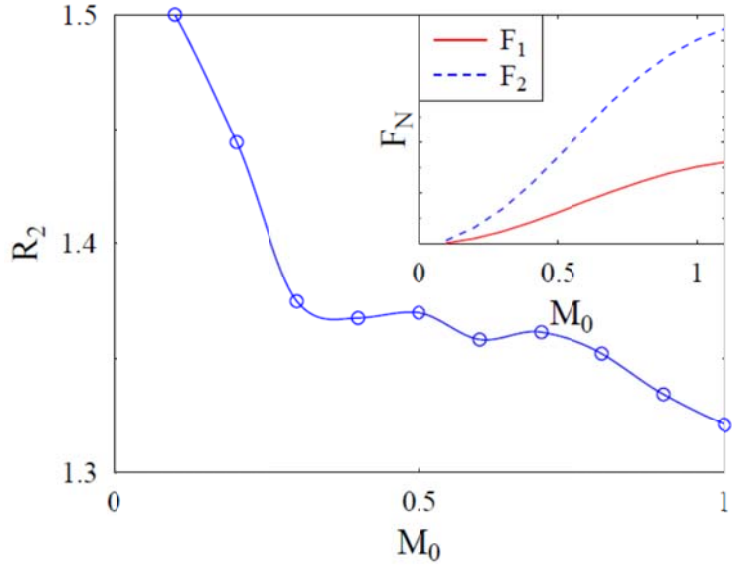

Fig. 4. The ratio $R_2$, evaluated at $E = E_{threshold}$ and used as a measure of cooperative inelastic scattering, plotted against the electron-vibration coupling parameter $M_0$. See text for the other parameters used in this calculation. The inset shows the inelastic signals $F_N(E=1)$ for $N=1,2$ in the same range of $M_0$.

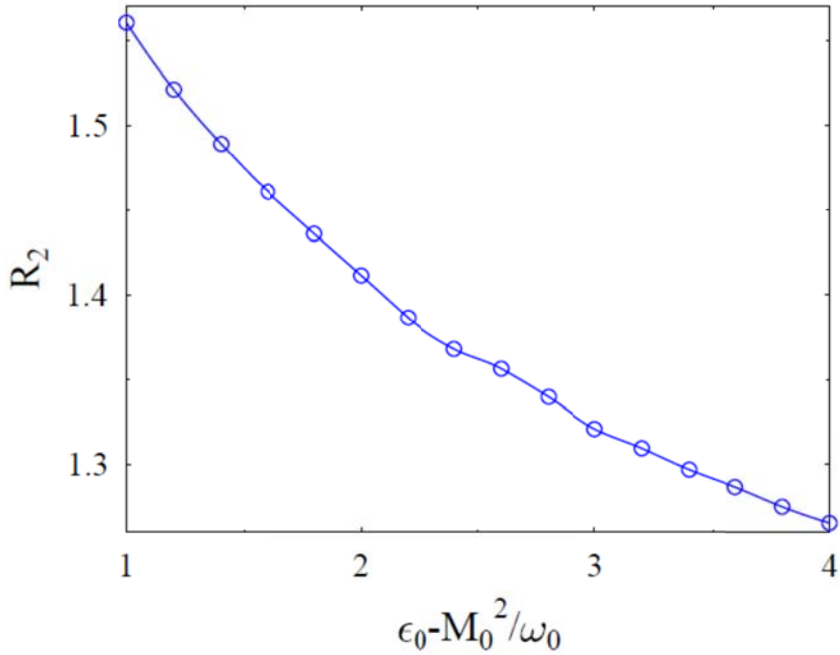

Fig. 5. The ratio $R_2(E = E_{threshold})$, plotted against the renormalized electronic resonance position, $\varepsilon_0 - M_0^2/\omega_0$. See text for other parameters.

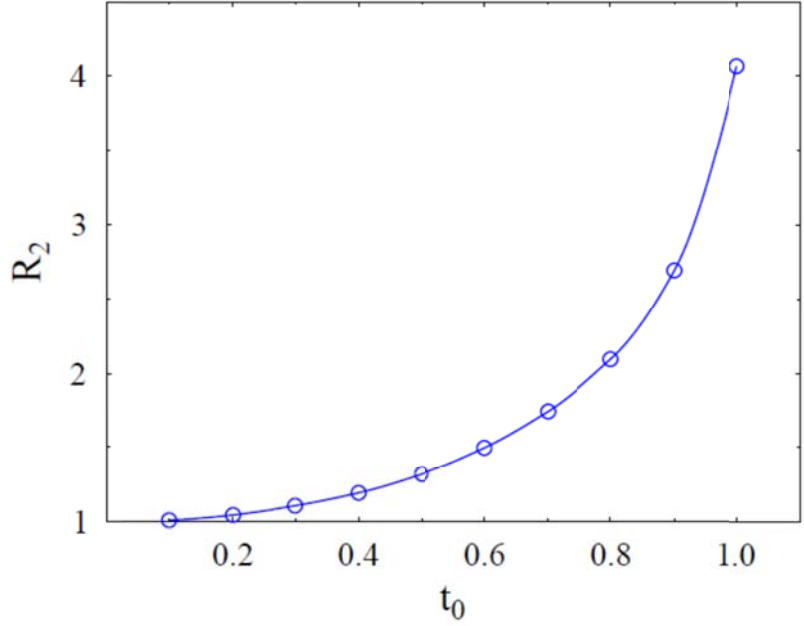

Fig. 6. The ration $R_2(E = E_{threshold})$ shown as a function of the renormalized electronic resonance position, $\varepsilon_0 - M_0^2/\omega_0$. See text for other parameters.

Figures 4-6 show the dependence of $R_N(E) \equiv F_N(E)/NF_1(E)$, $N = 2$, evaluated at the threshold energy $E = E_{threshold}$ on the electron-vibration coupling $M_0$, the renormalized molecular single electron energy $\bar{\varepsilon}_0 \equiv \varepsilon_0 - \dfrac{M_0^2}{\omega_0}$ and the molecule-metal coupling $t_0$. This ratio is used as our measure of cooperative inelastic tunneling in our model. The following observations are notable:

(a) The coherent (cooperative) inelastic response measure, $R$, decreases with increasing electron - vibration coupling $M_0$ (Fig. 4). Note that the inelastic signal itself (inset in Fig. 4) increases as expected with increasing $M_0$. A possible mechanism for this erasure of coherence is the participation of increasing number of vibration channels contributing to this signal with increasing coupling. These contributions may add up with different phases, leading to effective decoherence in the signal.

(b) The cooperative effect increases with the metal-molecule coupling $t_0$ as seen in Fig. 6. This is an expected behavior in our setup, where the molecular level is positioned above the band edge so that the tunneling process is of the super-exchange type. In fact, the resonance-position effect seen in Fig. 5, where the coherence decreases with increasing mismatch between the molecular electronic level and the metal band, is another manifestation of this phenomenon because the effective coupling for tunneling at energy $E(\sim E_{threshold})$ is $\sim t_{0L} t_{0R} / \Delta E$ where $\Delta E = E - \bar{\varepsilon}_0$.

## 5. Summary and conclusion

We have used a generic simple model to examine the possibility that coherent tunneling through a number of bridging molecules in a molecular junction may be accompanied by a cooperative inelastic response. As a measure of such cooperative response we have used the ratio $R_N = F_N / N F_1$ of the inelastic threshold peak heights in the IETS spectrum, $F = \left( d^2 I / dE^2 \right)_{E = E_{threshold}}$ of the $N$ molecules junction. We have used an extremely simple model that makes it possible to make an exact calculation by a generalized version of the Bonca-Trugman method. The fact that cooperative effects do exist in this model suggest that similar effects may show up in real molecular junctions and affect the analysis of observed inelastic spectra.

While we did not carry detailed calculations with a wide range of parameters, it is to be expected that cooperative effects in inelastic transport will become smaller for weak electron-vibration coupling, larger difference in timescales of vibrational and electronic dynamics (in the model considered here these timescales are characterized by the vibrational frequency and the electronic lead-molecular impurity coupling) and the presence of strong dephasing in the system. It is of interest to carry out such calculations, as well as similar calculations for a more realistic model of non-equilibrium (biased) molecular junction beyond the scattering theory based calculation done here. The latter calculation cannot be done by standard methods that disregards electron-vibration correlations in inelastic tunneling calculations, but could in principle be done using the pseudo-particle non-equilibrium Green function approached that was recently shown to account for such correlations.[45]

**Acknowledgements.** This paper is dedicated to the memory of Paul F. Barbara, a colleague, a friend and a forceful leading figure of our field. The research of A.N. is supported by the Israel Science Foundation, the Israel-US Binational Science Foundation, *the European Research Council under the European Union's Seventh Framework Program (FP7/2007-2013; ERC grant agreement n° 226628)* and the Israel – Niedersachsen Research Fund. M.G. gratefully acknowledges support by the NSF (Grant No. CHE-1057930) and the BSF (Grant No. 2008282).